\documentclass[11pt]{article} 

\usepackage{amsmath,amsthm}
\usepackage{mathtools}
\usepackage{microtype}

\usepackage{ifxetex,ifluatex}
\ifxetex
\usepackage{xltxtra}
\else
\ifluatex
\usepackage{luatextra}
\else
\usepackage{amssymb}
\usepackage[utf8]{inputenc}
\usepackage[T1]{fontenc}
 \usepackage{mathpazo,tgcursor,tgpagella}
 \usepackage{textcomp}
\mathtoolsset{mathic}

\usepackage[footnotesize,nooneline,bf]{caption}

\usepackage[mathscr]{eucal}

\newcommand{\BbbB}{\mathbb{B}}
\csname else\expandafter\endcsname
\romannumeral -`0
\fi
\fi
\iftrue\relax%
\defaultfontfeatures{Ligatures=TeX}
\usepackage[math-style=TeX, vargreek-shape=TeX]{unicode-math}
\fi

\usepackage[letterpaper,margin=1in]{geometry}

\usepackage{authblk}
\usepackage{enumitem}
\newlist{enumerate*}{enumerate*}{1}
\setlist[enumerate*]{label=(\arabic*),
  itemjoin={{, }}, itemjoin*={{, and }}}

\usepackage{xparse}

\usepackage[bookmarksnumbered,bookmarksopen,
unicode]{hyperref}

\hypersetup{
  pdftitle={An information diffusion Fano inequality},
  pdfauthor={Gábor Braun, Sebastian Pokutta},
  pdfkeywords={Rényi divergence, Fano inequality,
    continuous distribution},
  pdfsubject={MSC2000 Primary 94A17;
    Secondary 62B10}} 

\DeclareMathOperator{\supp}{supp} 
\newcommand{\R}{\mathbb{R}}
\newcommand{\face}[1]{\left\{#1\right\}}
\newcommand{\card}[1]{\left|#1\right|}
\newcommand{\vol}[1]{\operatorname{vol}(#1)}

\newcommand*{\VAR}[1]{\mathbf{#1}}
\newcommand*{\EVENT}[1]{\mathscr{#1}}

\NewDocumentCommand{\probability}{d()om}{%
  \operatorname{\mathbb{P}}%
  \IfValueT{#1}{\sb{#1}}%
  \left[#3%
    \IfValueT{#2}{\,\middle|\,#2}\right]}
\NewDocumentCommand{\expectation}{d()om}%
  {\operatorname{\mathbb{E}}%
      \IfValueT{#1}{\sb{#1}}%
      \left[#3%
    \IfValueT{#2}{\,\middle|\,#2}\right]}
\NewDocumentCommand{\entropy}{om}{\mathbb{H}\left[#2
    \IfValueT{#1}{\,\middle|\,#1}\right]}
\NewDocumentCommand{\bentropy}{lm}
  {\widetilde{\mathbb{H}}#1\left[#2\right]}
\newcommand*{\relEntropy}[2]
  {\operatorname{D}\left(#1\,\middle\|\,#2\right)}
\newcommand*{\brelEntropy}[2]
  {\operatorname{d}\left(#1\,\middle\|\,#2\right)}
\NewDocumentCommand{\mutualInfo}{omm}{\mathbb{I}\left[#2;#3
    \IfValueT{#1}{\,\middle|\,#1}\right]}

\NewDocumentCommand{\RenyiEntropy}{mom}{\mathbb{H}\sb{#1}\left[#3
    \IfValueT{#2}{\,\middle|\,#2}\right]}
\NewDocumentCommand{\bRenyiEntropy}{mlm}
  {\widetilde{\mathbb{H}}\sb{#1}#2\left[#3\right]}
\newcommand*{\RenyiDivergence}[3]
  {\operatorname{D}\sb{#1}\left(#2\,\middle\|\,#3\right)}
\newcommand*{\bRenyiDivergence}[3]
  {\operatorname{d}\sb{#1}\left(#2\,\middle\|\,#3\right)}

\DeclareMathOperator{\range}{range}

\theoremstyle{plain}
\newtheorem{theorem}{Theorem}[section]

\newtheorem{proposition}[theorem]{Proposition}

\newtheorem{corollary}[theorem]{Corollary}

\title{An information diffusion Fano inequality}

\author[1]{Gábor Braun}
\author[1]{Sebastian Pokutta}
\affil[1]{ISyE,
  Georgia Institute of Technology,
  Atlanta, GA 30332, USA.
  \textit{Email:}~ \{gabor.braun,sebastian.pokutta\}@isye.gatech.edu}

\begin{document}

\maketitle

\begin{abstract}
  In this note, we present an information diffusion inequality derived
  from an elementary argument, which
  gives rise to a very general Fano-type
  inequality. The latter unifies and generalizes the distance-based Fano inequality
  and the continuous Fano inequality established in \cite[Corollary~1,
  Propositions~1 and~2]{cont_Fano}, as well as the generalized Fano inequality
  in \cite[Equation following (10)]{han94_Fano}.
\end{abstract}

\section{Introduction}

Fano inequality is a crucial tool in information theory with
numerous applications. Moreover, it has been heavily used in
statistics in the context of minimax theory (see
\cite{lehmann1998theory} and references contained therein) and more recently
also in optimization (see e.g., \cite{raginsky2009information,agarwal2012information,bgp2013}) to
lower bound the rate of convergence of estimators and algorithms. The
general setup of Fano inequalities is a Markov chain \(\VAR X \to \VAR Y \to
\widehat{\VAR X}\) and we are interested in the probability of
finding a
sufficient \emph{reconstruction} \(\widehat{\VAR X}\) of the hidden random variable \(\VAR
X\) by observations \(\VAR Y\). Classically the measure of sufficiency
has been equality, i.e., we ask for perfect reconstructions
\(\widehat{\VAR X} = \VAR X\). This can be relaxed in several ways, by
e.g., accepting reconstructions \(\widehat{\VAR X}\), whenever \(\widehat{\VAR
  X}\) is close to \(\VAR X\).

In this note we present an elementary information diffusion
inequality, which immediately gives rise to a very general Fano
inequality, extending and subsuming the versions 
presented in \cite{cont_Fano}. In particular, we  allow for arbitrary relations
\(R \subseteq \range(\VAR X) \times \range(\widehat{\VAR X})\)
indicating a sufficient reconstruction.

Our notation is standard as to be found in \cite{cover2006elements},
and consistent with \cite{cont_Fano}.
We denote random variables by capital bold letters such as, e.g.,
\(\VAR X\) and events by scripts letters, such as \(\EVENT
R\).
Let \(\neg \EVENT{R}\) denote the negation of
the event \(\EVENT{R}\).

Let \(\log\) be a logarithm
with an arbitrary basis \(a > 1\),
which also serves as a basis for measuring information,
i.e., all information quantities are defined using
base \(a\) logarithm \(\log\).
Recall that the \emph{Rényi divergence}
of two distributions \(P\) and \(Q\)
over the same probability space is defined as
\begin{equation*}
  \RenyiDivergence{\alpha}{P}{Q}
  \coloneqq
  \frac{\log \expectation(P){
      \left(
        \frac{\mathrm{d} Q}{\mathrm{d} P}
      \right)^{1 - \alpha}}}{\alpha - 1}
\end{equation*}
for an order \(0 < \alpha < \infty\) with \(\alpha \neq 1\).
By continuity, this extends to orders \(0\), \(1\) and \(\infty\).
For the order \(\alpha = 1\) one recovers relative entropy,
also known as Kullback–Leibler divergence:
\begin{equation*}
  \RenyiDivergence{1}{P}{Q}
  =
  \relEntropy{P}{Q}
  \coloneqq
  \expectation(P){
      \log \left(
        \frac{\mathrm{d} P}{\mathrm{d} Q}
      \right)}
  .
\end{equation*}
When \(P\) and \(Q\) are Bernoulli distributions
with parameters \(p\) and \(q\) respectively,
we obtain the binary versions
\begin{align*}
  \bRenyiDivergence{\alpha}{p}{q}
  &
  \coloneqq
  \frac{\log \left(
    p^{\alpha} q^{1 - \alpha} +
    (1 - p)^{\alpha} (1 - q)^{1 - \alpha}
  \right)
  }{\alpha - 1}
  ,
  \\
  \brelEntropy{p}{q}
  &
  \coloneqq
  p \log \frac{p}{q}
  +
  (1 - p) \log \frac{1 - p}{1 - q}
  .
\end{align*}

The binary \emph{Rényi entropy} and binary \emph{entropy}
is defined as
\begin{align*}
  \bRenyiEntropy{\alpha}{p}
  &
  \coloneqq
  \frac{\log
    \left(
      p^{\alpha} + (1 - p)^{\alpha}
    \right)}{1 - \alpha}
  ,
  \\
  \bentropy{p}
  &
  \coloneqq
  p \log \frac{1}{p} + (1 - p) \log \frac{1}{1 - p}
  .
\end{align*}

\section{Information diffusion Fano inequality}
\label{sec:fano}

In this section we will present a general
information diffusion inequality, applicable to a broad range of
distributions, including continuous ones. We allow for 
specification of an arbitrary \emph{reconstruction relation} \(R \subseteq \range(\VAR X)
\times \range(\widehat{\VAR X})\), where \(\VAR X\) is a random variable
and \(\widehat {\VAR X}\) its reconstruction. We might want to think of \(R\) as specifying the acceptable
reconstructions, e.g., those with small
\(\ell_1\)-error. 

Our general Fano inequality is inspired by
a simple support-based lower bound on relative entropy,
see e.g., \cite[Theorem~3]{Renyi-div-cheatsheet}:
  For any two probability distributions \(P, Q\)
  on the same probability space,
  and denoting in the support \(\supp P\) of \(P\):
  \begin{equation*}
    \relEntropy{P}{Q} \geq \log \frac{1}{\probability(Q){\supp P}}.
  \end{equation*}

The next inequality is an extension of the generalized Fano
inequalities in \cite[Corollary~1 and Proposition~2]{cont_Fano},
where we do not consider the distance between \(P_{XY}\) and
\(P_X \times P_Y\) but rather between two arbitrary distributions
\(P_{XY}\) and \(Q_{XY}\). 

\begin{proposition}[Information diffusion Fano inequality]
  \label{prop:relEntropy-Fano}
  Let \(P\) and \(Q\) be two probability distributions
  on the same probability space and \(\EVENT{R}\) an event.
  Further, choose \(0 \leq p_{\min} < 1\) and \(0 < p_{\max} \leq 1\)
  with \(p_{\min} + p_{\max} < 1\) to be numbers
  satisfying
  \begin{equation}
    \label{eq:relEntropy-Fano-condition}
    p_{\min} \leq \probability(Q){\EVENT{R}} \leq p_{\max}.
  \end{equation}
  Then
  for any order \(0 < \alpha < \infty\) with \(\alpha \neq 1\):
  \begin{equation}
    \label{eq:RenyiDvergence-Fano}
    \probability(P){\EVENT{R}}
    \leq
    \sqrt[\alpha]{
      \frac
      {\exp\left[
          \left(
            \RenyiDivergence{\alpha}{P}{Q}
            +
            \bRenyiEntropy{\alpha}{\probability(P){\EVENT{R}}}
            +
            \log (1 - p_{\min})
          \right)
          \frac{\alpha - 1}{\log e}
        \right]
        - 1}
      {\left(
          \frac{1 - p_{\min}}{p_{\max}}
        \right)^{\alpha - 1}
        - 1}
    }
    .
  \end{equation}
  For the order \(\alpha = 1\),
  the following version holds:
  \begin{equation}
    \label{eq:relEntropy-Fano}
    \probability(P){\EVENT{R}} \leq
    \frac{\relEntropy{P}{Q} + \bentropy{\probability(P){\EVENT{R}}}
      + \log (1 - p_{\min})}
    {\log \frac{1 - p_{\min}}{p_{\max}}}
    .
  \end{equation}
\begin{proof}
The proof is an easy application of the data processing equality.
We shall also use the inequality
\begin{equation*}
  x^{\alpha} + y^{\alpha}
  \begin{cases}
    \geq (x + y)^{\alpha} & \text{if } \alpha \leq 1,
    \\
    \leq (x + y)^{\alpha} & \text{if } \alpha \geq 1,
  \end{cases}
  \qquad x, y > 0
\end{equation*}
with the choice \(x \coloneqq \probability(P){\EVENT{R}}\)
and \(y \coloneqq 1 - \probability(P){R}\):
\begin{equation}
  \label{eq:p-power-sum}
  \probability(P){R}^{\alpha} + (1 - \probability(P){R})^{\alpha}
  \begin{cases}
    \geq 1 & \text{if } \alpha \leq 1,
    \\
    \leq 1 & \text{if } \alpha \geq 1.
  \end{cases}
\end{equation}
One should verify the inequalities below separately
for \(\alpha < 1\) and \(\alpha > 1\).
\begin{align*}
  \RenyiDivergence{\alpha}{P}{Q}
  +
  \bRenyiEntropy{\alpha}{\probability(P){\EVENT{R}}}
  &
  \geq
    \bRenyiDivergence{\alpha}{\probability(P){\EVENT{R}}}
    {\probability(Q){\EVENT{R}}}
  +
  \bRenyiEntropy{\alpha}{\probability(P){\EVENT{R}}}
  \tag*{(data processing)}
  \\
  &
  =
  \frac{
     \log \left(
       \frac{
     {\probability(P){\EVENT{R}}}^{\alpha}
     {\probability(Q){\EVENT{R}}}^{1 - \alpha}
     +
     (1 - \probability(P){\EVENT{R}})^{\alpha}
     (1 - \probability(Q){\EVENT{R}})^{1 - \alpha}
   }
   {{\probability(P){\EVENT{R}}}^{\alpha}
     +
     (1 - \probability(P){\EVENT{R}})^{\alpha}}
     \right)
  }{\alpha - 1}
  \\
  &
  \tag*{(by Eq.~\eqref{eq:relEntropy-Fano-condition})}
  \geq
  \frac{
     \log \left(
       \frac{
     {\probability(P){\EVENT{R}}}^{\alpha}
     p_{\max}^{1 - \alpha}
     +
     (1 - \probability(P){\EVENT{R}})^{\alpha}
     (1 - p_{\min})^{1 - \alpha}
   }
   {{\probability(P){\EVENT{R}}}^{\alpha}
     +
     (1 - \probability(P){\EVENT{R}})^{\alpha}}
     \right)
  }{\alpha - 1}
  \\
  &
  =
  \frac{
    \log \left\{
      \frac{
        {\probability(P){\EVENT{R}}}^{\alpha}
        \left[
          \left(
            \frac{1 - p_{\min}}{p_{\max}}
          \right)^{\alpha - 1}
          - 1
        \right]
      }
      {{\probability(P){\EVENT{R}}}^{\alpha}
        +
        (1 - \probability(P){\EVENT{R}})^{\alpha}}
      + 1
    \right\}
  }{\alpha - 1}
  - \log (1 - p_{\min})
  \\
  &
  \tag*{(by Eq.~\eqref{eq:p-power-sum})}
  \geq
  \frac{
    \log \left\{
      {\probability(P){\EVENT{R}}}^{\alpha}
      \left[
        \left(
          \frac{1 - p_{\min}}{p_{\max}}
        \right)^{\alpha - 1}
        - 1
      \right]
      + 1
    \right\}
  }{\alpha - 1}
  - \log (1 - p_{\min})
   .
\end{align*}
The claim follows by rearranging.
For the case \(\alpha = 1\) we provide two proofs:
\begin{enumerate*}
\item by taking limit when \(\alpha \to 1\)
\item via a similar direct argument
\end{enumerate*}.
To simplify the limit argument,
let us introduce some shorthand notation:
\begin{align*}
  A_{\alpha}
  &
  \coloneqq
  \RenyiDivergence{\alpha}{P}{Q}
  +
  \bRenyiEntropy{\alpha}{\probability(P){\EVENT{R}}}
  +
  \log (1 - p_{\min})
  ,
  \\
  B
  &
  \coloneqq
  \frac{1 - p_{\min}}{p_{\max}}
  .
\end{align*}
Recall that
\(\lim_{\alpha \nearrow 1} \RenyiDivergence{\alpha}{P}{Q}
= \relEntropy{P}{Q}\),
therefore \(A_{1}\) is the numerator of \eqref{eq:relEntropy-Fano}.
The limit of the right-hand side of \eqref{eq:RenyiDvergence-Fano}
as \(\alpha \nearrow 1\)
\begin{equation*}
  \lim_{\alpha \nearrow 1}
  \sqrt[\alpha]{
    \frac
    {\exp\left[ A_{\alpha}
        \frac{\alpha - 1}{\log e}
      \right]
      - 1}
    {B^{\alpha - 1}- 1}
    }
  =
  \lim_{\alpha \nearrow 1}
  \sqrt[\alpha]{
    \frac{
      \frac{
        \exp\left( A_{\alpha}
          \frac{\alpha - 1}{\log e}
        \right)
        - 1}
      {A_{\alpha } \frac{\alpha - 1}{\log e}}}
    {\frac{B^{\alpha - 1} - 1}{\alpha - 1}}
    \cdot
    \frac{A_{\alpha}}{\log e}
  }
  =
  \sqrt[1]{
    \frac{1}
    {\frac{\log B}{\log e}}
    \cdot
    \frac{A_{1}}{\log e}
  }
  = \frac{A_{1}}{\log B},
\end{equation*}
which is exactly the right-hand side of
Eq.~\eqref{eq:relEntropy-Fano}.

An alternate proof via a direct computation goes as follows,
similar to the proof of Eq.~\eqref{eq:RenyiDvergence-Fano}:
\begin{align*}
   \relEntropy{P}{Q} + \bentropy{\probability(P){\EVENT{R}}}
   &
   \tag*{(data processing)}
   \geq
   \brelEntropy{\probability(P){\EVENT{R}}}
   {\probability(Q){\EVENT{R}}}
   + \bentropy{\probability(P){\EVENT{R}}}
   \\
   &
   =
   \begin{aligned}[t]
     &
     \probability(P){\EVENT{R}}
     \log \frac{\probability(P){\EVENT{R}}}
     {\probability(Q){\EVENT{R}}}
     +
     (1 - \probability(P){\EVENT{R}})
     \log
     \frac{1 - \probability(P){\EVENT{R}}}
     {1 - \probability(Q){\EVENT{R}}}
     \\
     &
     +
     \probability(P){\EVENT{R}}
     \log \frac{1}{\probability(P){\EVENT{R}}}
     +
     (1 - \probability(P){\EVENT{R}})
     \log
     \frac{1}{1 - \probability(P){\EVENT{R}}}
   \end{aligned}
   \\
   &
   =
   \probability(P){\EVENT{R}}
   \log \frac{1}{\probability(Q){\EVENT{R}}}
   + (1 - \probability(P){\EVENT{R}})
   \log \frac{1}{1 - \probability(Q){\EVENT{R}}}
   \\
   &
   \geq
   \probability(P){\EVENT{R}}
   \log \frac{1}{p_{\max}}
   + (1 - \probability(P){\EVENT{R}})
   \log \frac{1}{1 - p_{\min}}
   \tag*{(by~\eqref{eq:relEntropy-Fano-condition})}
   .
\end{align*}
Rearranging finishes the proof.
\end{proof}
\end{proposition}

We obtain a very general version of Fano's inequality as
a consequence.  This general form does not require any specific
distributional assumptions on \(\VAR X\) such as e.g., uniformity.
The case \(p_{\min} = 0\) is
\cite[Equation following (10)]{han94_Fano}.

\begin{proposition}[Fano inequality for arbitrary relations]
  \label{prop:Fano}
  Let \(\VAR X \to \VAR Y \to \widehat{\VAR X}\) be a Markov chain of
  random variables and let \(R\) be any set of values \((x, \widehat{x})\)
  with \(x \in \range(\VAR X)\) and \(\widehat{x} \in  \range(\widehat{\VAR X})\).
  Further, choose \(0 \leq p_{\min} < 1\), \(0 < p_{\max} \leq 1\)
  with \(p_{\min} + p_{\max} < 1\) to be numbers
  satisfying
  \begin{align*}
    p_{\min} \leq \inf_{\widehat{x}} \probability{(\VAR X, \widehat{x}) \in R}
    \qquad \text{and} \qquad  p_{\max} \geq \sup_{\widehat{x}}
    \probability{(\VAR X, \widehat{x}) \in R}.
  \end{align*}
  Let \(\EVENT{R}\) denote the event
  \((\VAR X, \widehat{\VAR X}) \in R\).
  Then
  \begin{equation}
    \label{eq:Fano}
    \probability{\EVENT{R}} \leq
    \frac{\mutualInfo{\VAR X}{\widehat {\VAR X}}
      + \bentropy{\probability{\EVENT{R}}} + \log (1 - p_{\min})}
    {\log \frac{1 - p_{\min}}{p_{\max}}}
\leq
    \frac{\mutualInfo{\VAR X}{\VAR Y} + \bentropy{\probability{\EVENT{R}}}
      + \log (1 - p_{\min})}
    {\log \frac{1 - p_{\min}}{p_{\max}}}
    .
  \end{equation}
\begin{proof}
The second inequality is equivalent to the data processing inequality
\(\mutualInfo{\VAR{X}}{\widehat{\VAR{X}}} \leq
\mutualInfo{\VAR{X}}{\VAR{Y}}\).
The first inequality is the following special case of
Proposition~\ref{prop:relEntropy-Fano}.
We choose \(P\) to be the joint distribution of
\((\VAR{X}, \widehat{\VAR{X}})\),
which is the distribution used in the statement,
i.e.,
\(\probability{\EVENT{R}} = \probability(P){\EVENT{R}}\).
We choose \(Q\) to be the product of the marginal distributions
of \(\VAR{X}\) and \(\widehat{\VAR{X}}\),
therefore \(\relEntropy{P}{Q} =
\mutualInfo{\VAR{X}}{\widehat{\VAR{X}}}\).
Finally,
\begin{equation*}
  \probability(Q){\EVENT{R}}
  =
  \probability(Q){(\VAR{X}, \VAR{\widehat{X}}) \in R}
  =
  \expectation(\widehat{x} \sim \widehat{\VAR{X}})
  {\probability{(\VAR{X}, \widehat{x}) \in R}}
  \geq
  \inf_{\widehat{x}} \probability{(\VAR X, \widehat{x}) \in R}
  \geq
  p_{\min},
\end{equation*}
and similarly,
\(\probability(Q){\EVENT{R}}
\leq  p_{\max}\).
Therefore the conditions of Proposition~\ref{prop:relEntropy-Fano}
are satisfied,
and its conclusion provides the first inequality
in \eqref{eq:Fano}.
\end{proof}
\end{proposition}

We immediately obtain the following corollary by
rearranging \eqref{eq:Fano}.
The condition \(p_{\min} + p_{\max} < 1\) is no longer needed, as
it was only used to preserve the direction of inequality
while dividing by \(\log [(1 - p_{\min}) / p_{\max}]\).
This step can be omitted by a direct proof, consisting of
repeating the last computation in the proof of
Proposition~\ref{prop:relEntropy-Fano}, and then rearranging.
\begin{corollary}[Entropy version of Fano inequality]\label{cor:entropyVersionFano}   Let \(\VAR X \to \VAR Y \to \widehat{\VAR X}\) be a Markov chain of
  random variables and let \(R\) be any set of values \((x, \widehat{x})\)
  with \(x \in \range(\VAR X)\) and \(\widehat{x} \in  \range(\widehat{\VAR
    X})\). With notation from Proposition~\ref{prop:Fano} we have
  \begin{align*}
     \entropy[\widehat {\VAR X}]{\VAR X} & \leq
    \entropy{\VAR X} + \log p_{\max} +
    \bentropy{\probability{\neg \EVENT{R}}}
    + \probability{\neg \EVENT{R}} \log \frac{1 - p_{\min}}{p_{\max}}
  \end{align*}
\end{corollary}

Moreover, if \(\VAR Y = (\VAR Y_1, \dots, \VAR Y_n)\) is obtained via
independent sampling from a hidden distribution specified by $\VAR X$,
i.e., the $\VAR Y_1, \dots, \VAR Y_n \mid \VAR X$ are i.i.d, then we
obtain the following corollary, which is sufficient for many
applications. The version with the relative entropy is obtained as a
direct consequence of the convexity of the relative entropy.

\begin{corollary}[Fano inequality for independent samples]
\label{cor:entropyVersionFanoSample}   Let \(\VAR X \to \VAR Y \to \widehat{\VAR X}\) be a Markov chain of
  random variables with \(\VAR Y = (\VAR Y_1, \dots, \VAR Y_n)\), so
  that $\VAR Y_1, \dots, \VAR Y_n \mid \VAR X$ are i.i.d. Further, let \(R\) be any set of values \((x, \widehat{x})\)
  with \(x \in \range(\VAR X)\) and \(\widehat{x} \in  \range(\widehat{\VAR
    X})\). With notation from Proposition~\ref{prop:Fano} we have
  \begin{equation}
    \label{eq:FanoSamp}
    \probability{\neg \EVENT{R}} \leq
    \frac{n \cdot \mutualInfo{\VAR X}{\VAR Y_1} + \bentropy{\probability{\EVENT{R}}} + \log (1 - p_{\min})}
    {\log \frac{1 - p_{\min}}{p_{\max}}}
    \leq
    \frac{n \cdot \beta + \bentropy{\probability{\EVENT{R}}} + \log (1 - p_{\min})}
    {\log \frac{1 - p_{\min}}{p_{\max}}}
    ,
  \end{equation}
  where $\beta = \max_{x,x' \in \range(\VAR X)} \relEntropy{\VAR Y_1 | \VAR X = x}{\VAR Y_1 | \VAR X = x'}$.
\end{corollary}

\subsection{Special cases}
\label{sec:special-cases}

We will now show how to obtain \cite[Corollary~1, Propositions~1
and~2]{cont_Fano} as special cases of the general Fano inequality
from above by choosing the relation \(R\) accordingly. 

\subsubsection*{Distance-based Fano inequality}
\label{sec:distance-based-fano}

 For the distance-based case, let \(\rho:
\range(\VAR X) \times \range(\VAR X) \rightarrow \R\) be a symmetric
function---typically a metric. Let \(\VAR X\) be a discrete random
variable with \(2 \leq \card{\range(\VAR X)} \leq \infty\). Furthermore let
\(\widehat{\VAR X}\) denote the reconstruction and assume
\(\range(\widehat{\VAR X}) = \range(\VAR X)\). For a given radius \(t\)
denote \(P_t \coloneqq \probability{\rho(\VAR X,\widehat{\VAR X}) >
  t}\). We then obtain as corollary, in the case where \(\VAR X\) is uniform:

\begin{corollary}(Distance-based Fano inequality \cite[Proposition
  1]{cont_Fano})
\label{cor:distFano}
  Let \(\VAR X \to \VAR Y \to \widehat{\VAR X}\) be a Markov chain of
  random variables with \(\VAR X\) uniform. For a given radius \(t \geq 0\) define
\[
N_t^{\max{}} \coloneqq \max_{x}\card{\face{\widehat{x} \mid \rho(x,\widehat{x}) \leq
  t}} \qquad \text{and} \qquad N_t^{\min{}} \coloneqq \min_{x} \card{\face{\widehat{x} \mid \rho(x,\widehat{x}) \leq
  t}},
\]
then 
  \begin{equation*}
\bentropy{P_t} + P_t \log \frac{\card{\range(\VAR X)} -
    N_t^{\min{}}}{N_t^{\max{}}} + \log N_t^{\max{}} \geq
\entropy[\widehat{\VAR X}]{\VAR X}
  \end{equation*}
\begin{proof}
We pick \(R \coloneqq \face{ (x,\widehat{x}) \in \range(\VAR X) \times
  \range(\VAR X) \mid \rho(x,\widehat{x}) \leq t}\),
so that \(\probability{\neg \EVENT{R}} = P_{t}\),
and choose
\( p_{\min} \coloneqq \frac{N_{t}^{\min}}{\card{\range(\VAR{X})}}\)
and
\( p_{\max} \coloneqq \frac{N_{t}^{\max}}{\card{\range(\VAR{X})}}\).
By Corollary~\ref{cor:entropyVersionFano}
using \(\entropy{\VAR{X}} \leq \log \card{\range(\VAR{X})}\)
\begin{equation*}
 \begin{split}
  \entropy[\widehat {\VAR X}]{\VAR X} & \leq
    \entropy{\VAR X} + \log
    \frac{N_{t}^{\max}}{\card{\range(\VAR{X})}}
    +
    \bentropy{P_{t}}  + P_{t} \log
    \frac{1 - \frac{N_{t}^{\min}}{\card{\range(\VAR{X})}}}
    {\frac{N_{t}^{\max}}{\card{\range(\VAR{X})}}}
    \\
    &
    \leq
    \log \card{\range(\VAR{X})} + \log
    \frac{N_{t}^{\max}}{\card{\range(\VAR{X})}}
    +
    \bentropy{P_{t}}  + P_{t} \log
    \frac{\card{\range(\VAR{X})} - N_{t}^{\min}}{N_{t}^{\max}}
    \\
    &
    =
    \log N_{t}^{\max}
    +
    \bentropy{P_{t}}  + P_{t} \log
    \frac{\card{\range(\VAR{X})} - N_{t}^{\min}}{N_{t}^{\max}}
    ,
 \end{split}
\end{equation*}
as claimed.
\end{proof}
\end{corollary}

Note that we require \(\VAR X\) to be uniform in
Corollary~\ref{cor:distFano} to easily match the form of
\cite[Proposition~1]{cont_Fano}. However, the uniformity requirement
can be removed. With the same choice for \(R\), we also immediately obtain
\cite[Corollary~1]{cont_Fano}, either by following the approach in
\cite{cont_Fano} or by directly invoking Proposition~\ref{prop:Fano}.

\begin{corollary}(Mutual information version of distance-based Fano inequality \cite[Proposition
2]{cont_Fano})   With the notation of Corollary~\ref{cor:distFano}, let \(\VAR X \to \VAR Y \to \widehat{\VAR X}\) be a Markov chain of
  random variables with \(\VAR X\) uniform.
  For any radius \(t \geq 0\) we have
\[ P_t \geq 1 - \frac{\mutualInfo{\VAR X}{\VAR Y} + \bentropy{P_t} }
    {\log \frac{\card{\range(\VAR X)}}{N_t^{\max}}}.\]
\end{corollary}

\subsubsection*{Continuous Fano inequality}
\label{sec:cont-fano-ineq}

In a next step, we will show how to
obtain the continuous Fano inequality of \cite{cont_Fano}, avoiding the discretization
argument altogether. Our version is slightly more general.

Let \(\VAR X\) be a continuos random variable so that that
\(\range(\VAR X)\) has finite non-zero Lebesgue measure. Moreover, let
\(\range(\widehat{\VAR X}) = \range(\VAR X)\) as in the discrete
distance-based setup. With the
notation from above, we define
\(\BbbB_\rho(t,x) \coloneqq \face{\widehat{x} \in \range(\VAR X) \mid
  \rho(x,\widehat{x}) \leq t}\). We obtain

\begin{corollary}(Continuos Fano inequality \cite[Proposition
2]{cont_Fano})
Let \(\VAR X \to \VAR Y \to \widehat{\VAR X}\) be a Markov chain of
  random variables with \(\VAR X\) uniform. For a given radius \(t \geq 0\) we have
\[ P_t \geq 1 - \frac{\mutualInfo{\VAR X}{\VAR Y} + \log 2}
    {\log \frac{\vol{\range(\VAR X)}}{\sup_{x}
        \vol{\BbbB_\rho(t,x) \cap \range(\VAR X) }}}.\]
\begin{proof}
As before, we choose \(R \coloneqq \face{ (x,\widehat{x}) \in \range(\VAR X) \times
  \range(\VAR X) \mid \rho(x,\widehat{x}) \leq t}\),
so that \(\probability(P){\EVENT{R}} = 1 - P_{t}\).
We apply Proposition~\ref{prop:Fano} with the choice
\(p_{\min} = 0\) and
\(p_{\max} = \frac{\sup_{x} \vol{\BbbB_\rho(t,x) \cap \range(\VAR X)}}
{\vol{\range(\VAR X)}}\)
and obtain
\begin{equation*}
  1 - P_{t}
  \leq
  \frac{\mutualInfo{\VAR X}{\VAR Y} + \bentropy{P_{t}}}
  {\log \frac{\vol{\range(\VAR X)}}
    {\sup_{x} \vol{\BbbB_\rho(t,x) \cap \range(\VAR X)}}},
\end{equation*}
which is the claim rearranged.
\end{proof}
\end{corollary}

\section*{Acknowledgements}
\label{sec:acknowledgements} 
Research reported in this paper was partially supported
by NSF grant CMMI-1300144 and CCF-1415496.

\bibliographystyle{alphaabbr}

\bibliography{gcNotesBib.bib}

\begin{thebibliography}{ABRW12}

\bibitem[ABRW12]{agarwal2012information}
A.~Agarwal, P.~L. Bartlett, P.~Ravikumar, and M.~J. Wainwright.
\newblock Information-theoretic lower bounds on the oracle complexity of
  stochastic convex optimization.
\newblock {\em Information Theory, IEEE Transactions on}, 58(5):3235--3249,
  2012.

\bibitem[BGP13]{bgp2013}
G.~Braun, C.~Guzm{\'a}n, and S.~Pokutta.
\newblock Unifying lower bounds on the oracle complexity of nonsmooth convex
  optimization.
\newblock {\em submitted}, 2013.

\bibitem[CT06]{cover2006elements}
T.~Cover and J.~Thomas.
\newblock {\em Elements of information theory}.
\newblock Wiley-interscience, 2006.

\bibitem[DW13]{cont_Fano}
J.~C. Duchi and M.~J. Wainwright.
\newblock Distance-based and continuum {F}ano inequalities with applications to
  statistical estimation.
\newblock {\em arXiv:1311.2669v2}, 2013.

\bibitem[HV94]{han94_Fano}
T.~S. Han and S.~Verd{\'u}.
\newblock Generalizing the {F}ano inequality.
\newblock {\em IEEE Transactions on Information Theory}, 40(4):1247--1251, July
  1994.

\bibitem[LC98]{lehmann1998theory}
E.~L. Lehmann and G.~Casella.
\newblock {\em Theory of point estimation}, volume~31.
\newblock Springer, 1998.

\bibitem[RR09]{raginsky2009information}
M.~Raginsky and A.~Rakhlin.
\newblock Information complexity of black-box convex optimization: A new look
  via feedback information theory.
\newblock In {\em Communication, Control, and Computing, 2009. Allerton 2009.
  47th Annual Allerton Conference on}, pages 803--510. IEEE, 2009.

\bibitem[vEH14]{Renyi-div-cheatsheet}
T.~van Erven and P.~Harremo{\"e}s.
\newblock R{\'e}nyi divergence and {K}ullback--{L}eibler divergence.
\newblock {\em arXiv:1206.2459v2 [cs.IT]}, April 2014.

\end{thebibliography}

\end{document}